\documentclass[aps,pre,twocolumn,showkeys,showpacs,amsmath,amssymb,groupedaddress]{revtex4-1}

\usepackage{graphicx}
\usepackage{bbm}
\usepackage[]{lineno}
\usepackage{color}
\usepackage{bm}

\begin{document}

\title{Sampling of Temporal Networks: Methods and Biases}

\author{Luis E C Rocha}\email{luis.rocha@ki.se}
\affiliation{Department of Public Health Sciences, Karolinska Institutet, Stockholm, Sweden}
\affiliation{Department of Mathematics, Universit\'e de Namur, Namur, Belgium}
\author{Naoki Masuda}
\affiliation{Department of Engineering Mathematics, University of Bristol, Bristol, UK}
\author{Petter Holme}
\affiliation{Institute of Innovative Research, Tokyo Institute of Technology, Tokyo, Japan}

\date{\today}

\begin{abstract}

Temporal networks have been increasingly used to model a diversity of systems that evolve in time; for example human contact structures over which dynamic processes such as epidemics take place. A fundamental aspect of real-life networks is that they are sampled within temporal and spatial frames. Furthermore, one might wish to subsample networks to reduce their size for better visualization or to perform computationally intensive simulations. The sampling method may affect the network structure and thus caution is necessary to generalize results based on samples. In this paper, we study four sampling strategies applied to a variety of real-life temporal networks. We quantify the biases generated by each sampling strategy on a number of relevant statistics such as link activity, temporal paths and epidemic spread. We find that some biases are common in a variety of networks and statistics, but one strategy, uniform sampling of nodes, shows improved performance in most scenarios. Our results help researchers to better design network data collection protocols and to understand the limitations of sampled temporal network data.
\end{abstract}


\maketitle

\section{Introduction}
\noindent

Networks have been used to model the interactions and interdependencies between the parts of a system~\cite{Newman2010}. Social and sexual contacts, flights between airports, email and phone communication, or gene regulatory networks are just a few examples of systems that can be conveniently mapped into networks~\cite{Newman2010, Costa2011}. When modelling real-systems as networks, researchers sample data by extracting the relevant information within a given temporal and spatial frame~\cite{Lohr2009}, trace-routing or snow-balling from one or multiple sources~\cite{Sudman1988, Achlioptas2008}, or simply by collecting all network-related information of a specific system, for example, email exchanges within a university or social interactions on a web-community~\cite{Lee2006, Costa2011}. Sampling network data involves at least four main decisions: the choice of (i) the total observation, or sampling, time (e.g.\ 1 day or 1 year); (ii) which nodes and (iii) links will be observed (e.g.\ all or a fraction); (iv) the temporal resolution, i.e.\ the time interval in which data are recorded. If the temporal resolution is smaller than the total observation time, several interaction events between the same pair of nodes may be recorded and filtering strategies may be used to remove weak links~\cite{Serrano2009}.

Network modelling may involve the traditional framework of static networks or extensions such as temporal networks~\cite{Holme2015}. In temporal networks, nodes and links are active at given times in contrast to static networks where nodes and links remain active during the whole period. Temporal networks thus describe more realistically the temporal paths through which information (e.g.\ through email communication~\cite{Eckmann04}), infections (e.g.\ over sexual contacts~\cite{Masuda2013}), or resources (e.g.\ via flights~\cite{Rocha2016b}) can propagate or flow. In this temporal perspective, the order and frequency of node and link activations directly affect the dynamics of simulated epidemics~\cite{riolo_etal,fffng,Rocha2011,Stehle2011, Lambiotte13, Speidel2016} and information spread~\cite{lamport, moody, Vazquez07, Karsai2011PhysRevE}, mixing properties of random walks~\cite{Lambiotte13, Scholtes2014, Delvenne2015}, and synchronization~\cite{Masuda13, Masuda2016} on networks. Although some level of recording error is acceptable, accurate labeling of the interaction events is important to study, for example, simulated infections on real-life temporal networks~\cite{Holme2017}.

Another challenge that comes with the study of real-life temporal networks is the amount of generated data since all timings of link activation are stored. This is in contrast to static networks in which activation events are aggregated and multiple activations of the same link are then represented as single weights~\cite{krings}, saving memory. The memory cost is particularly problematic when handling big data or when designing studies to collect social interactions using electronic devices such as RFID tags~\cite{Barrat2013,thebook:barrat} and mobile phones~\cite{stopczynski2014measuring}. In both cases, researchers aim to collect as much relevant data as possible while optimising resources. Furthermore, several algorithms used to extract information or to simulate dynamic processes on networks struggle to deal with large temporal networks, becoming computationally intractable~\cite{Pan2011, Gauvin2014, Vestergaard2015, Rocha2016, Paranjape2017}. Facing these challenges, the natural question that emerges is what data should be collected and used in network studies.

The four sampling decisions mentioned above are more critical for the study of temporal networks than for static networks. For example, the total observation time might affect birth and death statistics of nodes and links, and add artificial cutoffs to inter-event times since interaction events might be truncated. Similarly, the temporal resolution acts like a filter since only temporal patterns of node and link activity at time scales above the resolution are observable. A typical example is to use a resolution of 1 day to collect data on email communication; this choice would miss the rich dynamic communication patterns happening within a day. The sampling of nodes and links are expected to have at least the same effect as on static networks~\cite{Lee2006, Achlioptas2008} with the aggravated consequence that missing nodes and links would also affect the temporal patterns of the neighbouring nodes.

When sampling temporal network data, one wishes to collect as much information as possible such that both short- and long-term temporal patterns can be observed~\cite{saramoro}. Yet, the amount of information should be manageable by existing algorithms. In this paper, we study the impact of four sampling design decisions, or strategies, on key temporal network variables applied to various categories of real-life temporal networks. In particular, we will study how the choice of the observation time, the temporal resolution, and the number of sampled nodes and events, affect the statistics of the lifetime and burstiness of links, the number and length of temporal paths between nodes, and the number of secondary infections and outbreak size of simulated epidemics.

\section{Materials and Methods}

\subsection{Temporal Networks}

For a given time period $T$, a temporal network of size $N$ is defined as a set of nodes $i$ connected by a set $E$ of links $(i,j)$, in which events occur at times $t$~\cite{Holme2015}. $M$ represents the sum of the number of events in each link for all links. The temporal resolution $\delta$ characterises the size of the time interval (or snapshot) in which the network data are collected, therefore, an event occurring at time $t$ actually means that the event occurred in the time interval $[t,t+\delta)$. The statistics of the sampled networks will be represented by the subscript $s$, for example, $N_s$ and $M_s$ represent the number of nodes and of events, respectively.

\subsection{Network Data}

We will use six network data sets corresponding to different contexts in which temporal networks are relevant. We have chosen networks with different topological and temporal structures. The first data set corresponds to sexual contacts between sex-workers and their clients (SEX)~\cite{Rocha2010,Rocha2011}; the second is about online communication between users of a web-community related to movies (FOR)~\cite{Karimi2014}; the third is about email communication within a university (EMA)~\cite{Eckmann04}; the fourth is about online communication between students in an online social network (COL)~\cite{Panzarasa2009}; the fifth is about face-to-face proximity contacts ($\le 1.5$m) between high-school students (HSC)~\cite{Mastrandrea2015}; the sixth is also about proximity contacts but between visitors of a museum exposition (GAL)~\cite{Broeck2012} (See Table~\ref{tab:01}). Links are undirected and only a single event may occur in a time window $[t,t+\delta)$ for a given link, i.e.\ events are unweighted.

\begin{table}[htb]
\begin{center}
\caption{Summary statistics of the original temporal networks. Number of nodes ($N$), number of events ($M$), temporal resolution ($\delta$), and observation time ($T$).}
\begin{tabular}{cccccc}
\hline
       & $N$ & $M$ & $\delta$ & $T$ \\
\hline
SEX & 11,416 & 33,508  & 1 day & 1,000 days   \\
FOR & 7,084  & 625,435 & 1 day & 3,142 days    \\
EMA & 3,186  & 234,412 & 1 hour & 1,959 hours \\
COL & 1,899  & 37,178   & 1 hour & 4,649 hours  \\
HSC & 310     & 47,338   & 20 sec &  32,360 sec  \\
GAL & 204     & 6,709     & 20 sec & 29,000 sec   \\
\hline
\end{tabular}
\label{tab:01}
\end{center}
\end{table}

\subsection{Sampling Methods}

Sampling consists in making a number of observations or selecting a set of individuals to estimate properties of the target population. In the context of networks, sampling means selecting a number of nodes and links of a system within temporal and spatial frames to build the network of interest. In this paper, we will take network data sets available in the literature as reference populations. We will then study the consequences, on the network structures and on dynamics on the networks, of applying different sampling strategies on these populations. Effectively, we will subsample the original empirical network and then discuss the biased estimates of each sample, that is, the difference in the estimates given by the sampled and the original networks. This subsampling approach is widely used in statistics (see e.g.\ subsampling bootstrap~\cite{Politis1999}) and other disciplines (see e.g.\ \cite{Lee2006, Achlioptas2008}).

\begin{figure}[thb]
\centering
\includegraphics[scale=1]{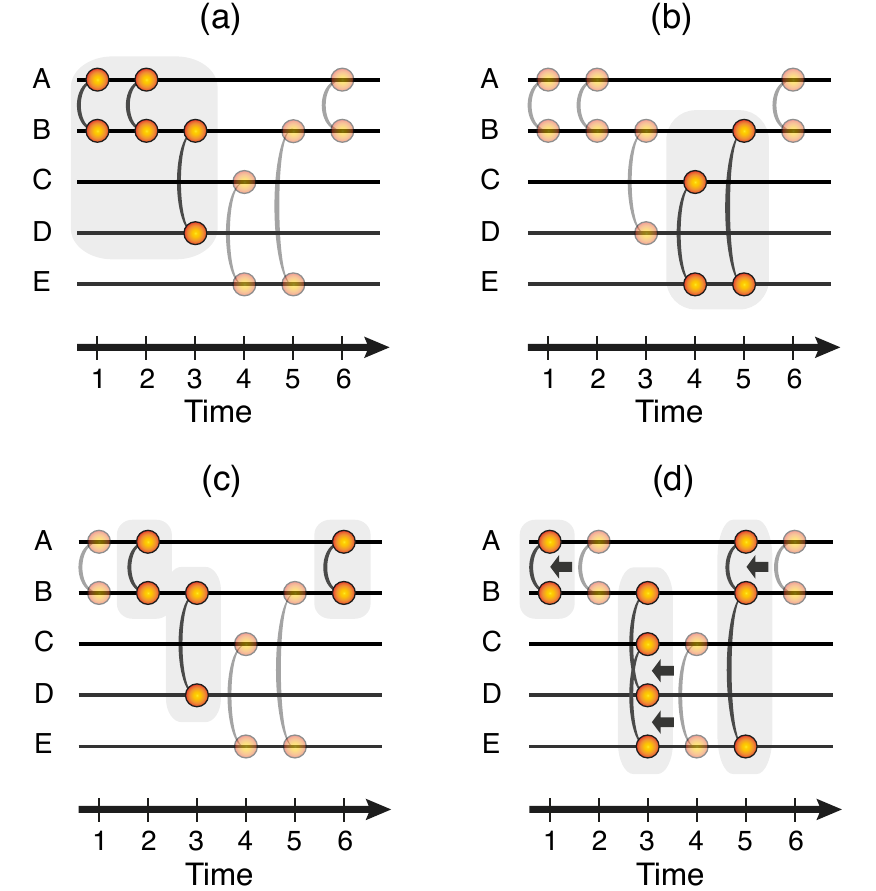}
\caption{\textbf{Sampling strategies.} All panels show a time-line representation of a temporal network where one horizontal line represents a node and there is a vertical line connecting two nodes if they interact at a particular time (i.e.\ an event). Sampled nodes and sampled events are highlighted for each strategy. In (a), we obtain a new temporal network by truncating the observation time to $T_s$. In this example $T_s=3$, therefore all nodes and events in $1\le t \le 3$ are collected. In (b), we uniformly choose nodes in $1 \le t \le T$. In this example, nodes $B$, $C$ and $E$ are sampled, therefore only the events between these particular nodes are collected. In (c), we uniformly choose events in $1 \le t \le T$. In this example, the events $(A,B)$ at $t=2$, $(B,D)$ at $t=3$, and $(A,B)$ at $t=6$ are sampled. In (d), we coarse-grain the temporal network in the interval $1 \le t \le T$ by letting an event represent the presence of at least one event at that link during $\delta_s$. In this example, we change the resolution from $\delta = 1$ to $\delta_s = 2$, therefore we only record interaction events at times 1, 3, and 5, and events are merged if they repeat (e.g.\ original events at $t=1$ and $t=2$ at link $(A,B)$ become a single event at $t=1$). }
\label{fig:01}
\end{figure}

We will study the effect of four sampling strategies (Fig.~\ref{fig:01}): (i) to reduce the observation time $T_s$, where $T_s \leq T$ and $[0,T_s]$ is the sampling time in which the network data are collected (strategy TS); (ii) to uniformly select a fraction $N_s/N$ of nodes of the original network and thus all events between the sampled nodes (strategy NS); (iii) to uniformly select a fraction $M_s/M$ of events of the original network and thus all nodes connected by these events (strategy ES) -- note that this protocol is used, instead of selecting links (and consequently all events associated to that particular link), because of higher flexibility and because one can design ``on-line sampling'', that is, collect events as they happen in time; and (iv) to reduce the resolution by setting $\delta_s$ a multiple of $\delta$ of the original network (strategy RS). Note that repeated same-link events in the interval $[t,t+\delta_s)$ are merged into a single event.

\subsection{Validation Measures}
\label{sec:val_meas}

To compare the effects of the four sampling strategies, we will estimate six measures, or statistics, on each sample $s$ of the original networks. For strategies NS and ES, we will present average values calculated over five random network samples. Two measures are related to the timings of events, two to the temporal paths and two to the dynamics on the network.

The first measure is the burstiness $B_s$ of the link activity~\cite{Goh2008}. This measure is widely used to characterise temporal patterns on temporal networks. The burstiness depends on the mean $m$ and standard deviation $\sigma$ of the distribution of same-link inter-event times (the inter-event time is the time between two subsequent same-link activations) and measures the deviation of the link activity from a Poisson process. Considering the distribution of inter-event times of all links collected together, the burstiness is given by
\begin{equation}
\label{eq:01}
B_s = \frac{\sigma - m}{\sigma + m}.
\end{equation}

The second measure is related to the lifetime $L_{ij}$ (or persistence~\cite{clauset_eagle}) of links, that is, the time between the first event, $t^{\text{first}}_{ij}$, and last event, $t^{\text{last}}_{ij}$, on the link $(i,j)$.
The link lifetime can be used as a proxy for the real lifetime of contacts~\cite{ongoing}. We measure the average lifetime $L_s$ over all $K_s$ links in which $L_{ij}>0$ (i.e.\ there are at least two events in the link) to summarise the lifetime of the links in the sampled network, i.e.
\begin{equation}
\label{eq:02}
L_s = \frac{1}{K_s} \sum_{(i,j)\in E_s, L_{ij}>0} \left( t_{ij}^{\text{last}} - t_{ij}^{\text{first}} \right).
\end{equation}

The third and fourth measures are related to temporal paths. Temporal paths are particularly relevant in the context of temporal networks because they combine topological and temporal information. They emphasise the role of the timings of events in the connectivity of a node. For example, two nodes may be topologically close (e.g.\ directly connected by a link) but one may need to wait a long time for this link to be active (i.e.\ for an interaction event to happen). On the other hand, a more topologically distant pair of nodes (e.g.\ two links away) may be reached quickly if the interaction events are temporally close. We assume here that, within a time step, a node can only be reached by another node through a direct link. For example, there are no paths connecting nodes A and C if the events (A,B) and (B,C) occur at the same time. An alternative assumption could define a path between A and C in this example~\cite{Tang2010}.

The third measure is the reachability ratio $f_s$~\cite{Holme2005}. It is the fraction of pairs of nodes that have at least one temporal path between them and is defined by
\begin{equation} 
\label{eq:03}
f_s = \frac{1}{N_s(N_s-1)} \sum_{i,j=1}^{N_s} {\mathbbm{1}(\tau_{ij})},
\end{equation}
where:
\[ \mathbbm{1}(\tau_{ij}) =
  \begin{cases}
    1 & \quad \text{if } \tau_{ij} \text{ exists,}\\
    0 & \quad \text{ otherwise.}\\
  \end{cases}
\]

It can happen that $\tau_{ij}$ is finite, whereas $\tau_{ji}$ is infinite, or vice versa.

The fourth measure is related to the time distance between nodes in the network~\cite{Holme2005, Pan2011, Masuda2016b}. The time distance $\tau_{ij}$ is here defined as the time necessary to reach node $j$ from the first appearance (i.e.\ birth) of node $i$ through the shortest temporal path connecting $i$ and $j$. If there is no path between nodes $i$ and $j$, we set $\tau_{ij} \rightarrow \infty$~\cite{Masuda2016b}. We then set

\begin{equation} 
\label{eq:04}
\theta_s = \frac{1}{N_s(N_s-1)} \sum_{i,j =1}^{N_s} \frac{1}{\tau_{ij}}.
\end{equation}
to summarise $\tau_{ij}$ over the links. Note that $\tau_{ij} \rightarrow \infty$ contributes zero to the sum in Eq.~\eqref{eq:03} and that both the shortest path from $i$ to $j$ and that from $j$ to $i$ appear in Eq.~\eqref{eq:03} because $\tau_{ij}$ is not equal to $\tau_{ji}$ in general. This measure is normalized by $N_s(N_s-1)$, that gives the total number of possible paths between any two pairs of nodes if all links occur at the same time~\cite{Tang2010}.

For the fifth and sixth measures, we model a susceptible-infected-recovered (SIR) epidemics on the temporal network. In the SIR model, a node can be either susceptible (S), infected (I) or recovered (R). Infected nodes can infect susceptible nodes with probability $\beta$ and recover with probability $\mu$ in a time step. For strategy RS, to account for the change in the resolution $\delta_s$ (and consequently in the contact rate), we re-scale the parameters to $\beta/\delta_s$ and $\mu/\delta_s$. Re-scaling these parameters effectively conserves the contact rate because we assume the events are unweighted; without the re-scaling, the infection and recovery probabilities would be over-estimated for $\delta_s > \delta$. We start by infecting a single node and leaving all others susceptible. Under the so-called individual-based approximation~\cite{Rocha2016}, the dynamics of the probability that node $i$ is infected at time $t$ is given by
\begin{align}
\label{eq:05}
S_i(t) =& S_i(t-1) \prod_{j\in {{\cal N}_s}_i(t)} \phi_j(t), \\
I_i(t) =& I_i(t-1) + S_i(t-1) \left[ 1 - \prod_{j\in {{\cal N}_s}_i(t)} \phi_j(t) \right] - \mu I_i(t-1), \\
R_i(t) =& 1 - S_i(t) - I_i(t),
\end{align}
where ${{\cal N}_s}_i(t)$ is the set of neighbors of node $i$ at time $t$,
$\phi_j(t) = 1 -  (1-\mu)\beta I_j(t-1)$ if there is an event between nodes $i$ and $j$ at time $t$, and $\phi_j(t) = 1$ otherwise.

We then measure the average number of secondary infections $R^{\text{eff}}_s$ and the average final outbreak size $\Omega_s$ caused by a single infected node at time 0 for each sampled network~\cite{Rocha2016}. $R^{\text{eff}}_s$ is thought to indicate the propensity of an outbreak to become pandemic~\cite{Dietz1993}. The value of $\Omega_s$ is not linearly related to $R^{\text{eff}}_s$ although a larger $\Omega_s$ is expected for larger $R^{\text{eff}}_s$~\cite{Holme2015b}. Under the individual-based approximation, we obtain
\begin{equation}
\label{eq:06}
R^{\text{eff}}_s = \frac{1}{N_s} \sum_{i=1}^{N_s} \left[ \sum_{t=1}^{T_s} \left[1-\phi_i(t)\right]\sum_{j\in {{\cal N}_s}_i(t)}S_j(t-1) \right].
\end{equation}
and
\begin{equation}
\label{eq:07}
\Omega_s = \frac{1}{N_s} \sum_{i=1}^{N_s} \left[ I_i(T_s) + R_i(T_s) \right].
\end{equation}

\begin{figure}[thb]
\centering
\includegraphics[scale=1.0]{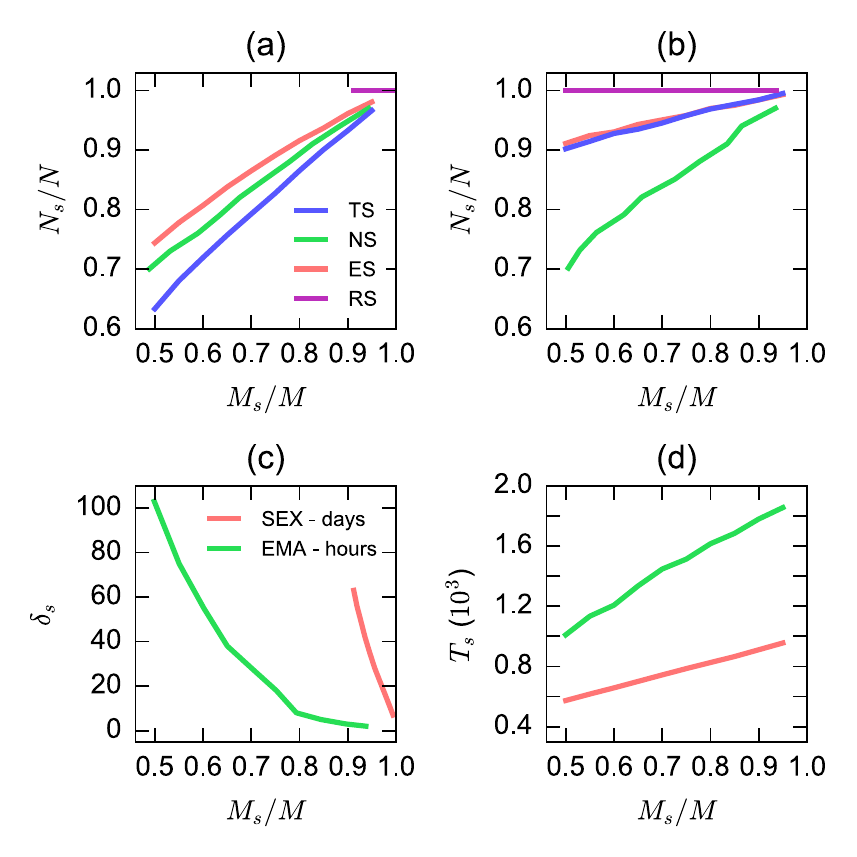}
\caption{\textbf{Characteristics of sampled networks.} The fraction of nodes ($N_s/N$) and events ($M_s/M$) after sampling the original networks using each of the sampling strategies for the (a) SEX and (b) EMA networks. In (c), we show the fraction of events ($M_s/M$) for a given temporal resolution ($\delta_s$). In (d), we show the fraction of events ($M_s/M$) for a given observation time $(T_s)$. Vertical bars for strategies NS and ES correspond to the standard deviation that is only visible if larger than the thickness of the curves.}
\label{fig:02}
\end{figure}

\section{Results}


\subsection{Network size}
\label{sec:net_size}
Different sampling strategies have a different impact on the number of nodes and events in the sampled networks (Fig.~\ref{fig:02}a,b). Reducing the temporal resolution $\delta_s$ (strategy RS) has no effect on the number of nodes ($N_s$) but monotonically decreases the number of events ($M_s$). This happens because some events repeat at subsequent times. If there is little repetition, reducing the temporal resolution will only slightly decrease the number of events. In the SEX network ($\delta=1$ day), for example, setting $\delta_s=63$ days only reduces the number of events by $8.87\%$ (Fig.~\ref{fig:02}a,c). This is the reason for the short magenta curve in Fig.~\ref{fig:02}a. In contrast, setting $\delta_s=55$ hours in the EMA network ($\delta=1$ hour) reduces the number of events by about $40\%$ (Fig.~\ref{fig:02}b,c). The high turnover of nodes (i.e.\ shorter lifetimes in comparison to the observation time in the original network) in the SEX network explains why the number of nodes falls more substantially in this case than in the EMA network if we reduce $T_s$ (strategy TS). For example, a reduction of about $43\%$ in $T_s$ results in about $37\%$ less nodes in the SEX network (Fig.~\ref{fig:02}a). For the EMA network however the reduction of $48\%$ in $T_s$ implies on only $9.8\%$ less nodes (Fig.~\ref{fig:02}b). The same reduction in $T_s$ by half results in approximately half the events in both cases (Fig.~\ref{fig:02}d).

The uniform sampling of events (strategy ES) has less impact on the number of nodes than the uniform sampling of nodes (strategy NS) if we control for the number of events (Fig.~\ref{fig:02}a,b). This happens because a node typically has more than one event with the same or with different neighbours. In strategy ES, highly connected nodes are selected often (proportionally to the number of events~\cite{Feld1991}) and thus sampled nodes might repeat, decreasing the final number of nodes in the sample. In strategy NS, on the other hand, the selection of nodes brings all their events (to other sampled nodes), implying that less nodes are selected (in comparison to strategy ES) for the same number of events.

In the following analyses, we will present the results for COL, FOR, HSC and GAL using two configurations ($A$ and $B$) for each strategy. Each configuration corresponds to a fixed number of events $M_s$. $M_s$ was based on an arbitrarily chosen resolution. That is, we set a resolution $\delta_s$ and took the number of events of this sample as reference to be used in the other sampling strategies. For the COL data set, $A$ corresponds to a fraction of $62\%$ ($\delta_s=48$ hours) and $B$ to a fraction of $77\%$ ($\delta_s=12$ hours) of the events of the original network. For the FOR data set, we have respectively $56\%$ ($\delta_s=24$ hours) and $74\%$ ($\delta_s=6$ hours), for HSC, $54\%$ ($\delta_s=60$ sec) and $68\%$ ($\delta_s=40$ sec), and for the GAL data set, we have $57\%$ ($\delta_s=60$ sec) and $70\%$ ($\delta_s=40$ sec).

\subsection{Timings of events}
\label{resul:timings}

\begin{figure}[thb]
\centering
\includegraphics[scale=1.0]{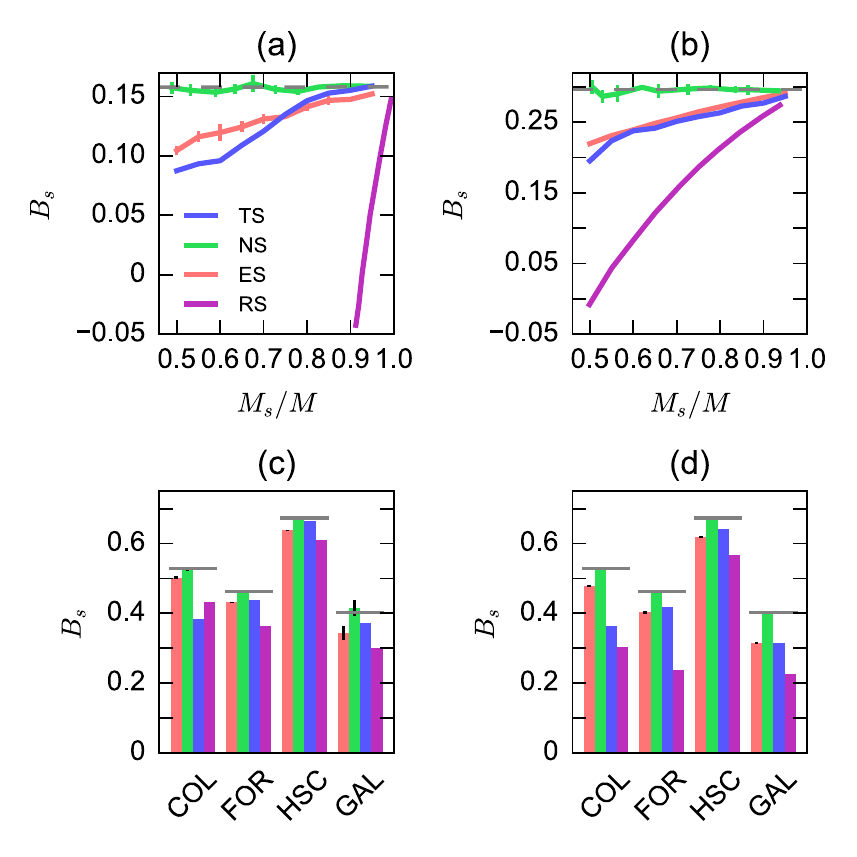}
\caption{\textbf{Burstiness.} The burstiness ($B_s$) of link activity (events) after sampling the original network using each of the sampling strategies for the (a) SEX and (b) EMA networks. The estimation of $B_s$ for the COL, FOR, HSC and GAL networks with configurations (c) $A$ and (d) $B$ (see Section~\ref{sec:net_size}). Dashed horizontal lines correspond to the results for the original networks. Vertical bars for strategies NS and ES correspond to the standard deviation that is only visible if larger than the thickness of the curves.}
\label{fig:03}
\end{figure}

We have found that uniformly sampling nodes (strategy NS) seems to be the best strategy to conserve the burstiness. The value of $B_s$ is robust in both SEX and EMA data sets even when only half of the events are sampled (Fig.~\ref{fig:03}a,b). The fact that the number of sampled nodes (by strategy NS) has little impact on the estimation of the burstiness suggests that all nodes follow similar inter-event times distributions, (i.e.\ a few nodes are sufficient for an accurate estimation, Fig~\ref{fig:02}a,b). On the other hand, increasing $\delta_s$ (strategy RS) has a significant negative effect on $B_s$. The resolution affects the distribution of inter-event times since increasing $\delta_s$ filters out short inter-event times and reduces the long inter-event times, making the signal move towards more regularity (with larger mean and standard deviation). Strategies ES and TS also generate biases, which are considerably smaller than biases given by strategy RS. For different reasons, strategies ES and TS also affect the distribution of inter-event times but to a lesser extent than strategy RS. Strategy ES misses a few events and thus increases the average (and standard deviation of the) inter-event times. In contrast, strategy TS skips events that could generate long inter-event times since the observation time is truncated and thus generates smaller means and standard deviations. Similar results are observed for the other data sets (Fig.~\ref{fig:03}c,d).

\begin{figure}[thb]
\centering
\includegraphics[scale=1.0]{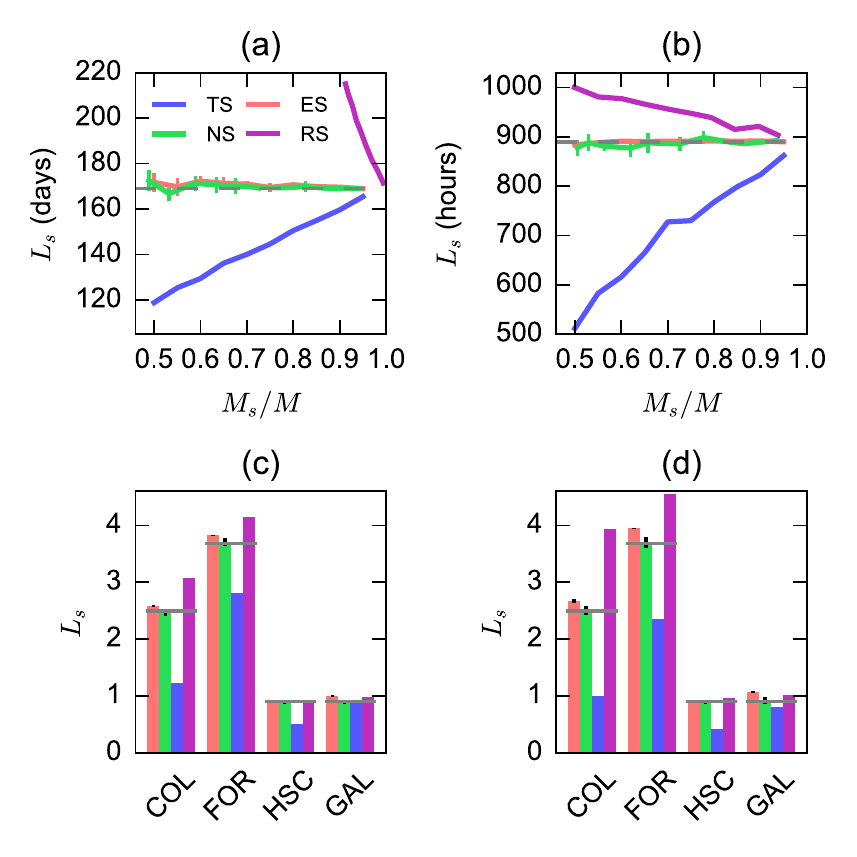}
\caption{\textbf{Lifetime.} The average lifetime ($L_s$) of links after sampling the original network using each of the sampling strategies for the (a) SEX and (b) EMA networks. The estimation of $L_s$ for the COL ($\times 10^2$ hours), FOR ($\times 10^2$ days), HSC ($\times 10^4$ seconds) and GAL ($\times 10^3$ seconds) networks with configurations (c) $A$ and (d) $B$. Dashed horizontal lines correspond to the results for the original networks. Vertical bars for strategies NS and ES correspond to the standard deviation.}
\label{fig:04}
\end{figure}

Strategies NS and ES generally give good estimations of the average lifetime of links $L_s$ for all data sets (Fig.~\ref{fig:04}a-d). The uniform sampling of events or nodes decreases the lifetime of some links but also sometimes does not sample any event of a particular link (i.e.\ some links and nodes may not be sampled at all). The smaller $K_s$ possibly compensates the decrease in the lifetimes such that the average $L_s$ is little affected. Strategy TS introduces cut-offs on the lifetimes of both links and nodes since sampling is limited within the observation time $[0,T_s]$. Consequently, the lifetime is underestimated. The case of GAL is special because visitors explore the museum in groups at allocated times, meaning that links form and disappear before $T_s$ (Fig.~\ref{fig:04}c,d). Finally, strategy RS tends to overestimate $L_s$ because increasing $\delta_s$ is equivalent to rounding down the times of births and deaths. The rounding down leads to an overall increase in the lifetime of links and a decrease in $K_s$ since links with a single event are not included in the average.

\subsection{Temporal Paths}

\begin{figure}[thb]
\centering
\includegraphics[scale=1.0]{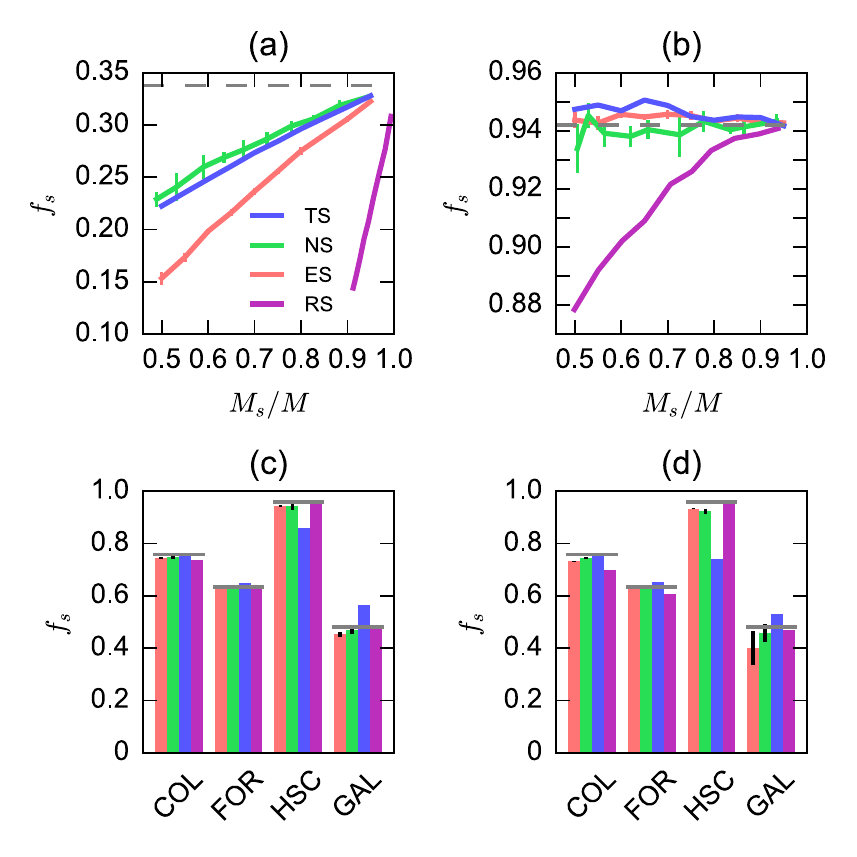}
\caption{\textbf{Number of temporal paths.} The fraction of temporal paths ($f_s$) after sampling the original network using each of the sampling strategies for the (a) SEX and (b) EMA networks. The estimation of $f_s$ for the COL, FOR, HSC and GAL networks with configurations (c) $A$ and (d) $B$. Dashed horizontal lines correspond to the results for the original networks. Vertical bars for strategies NS and ES correspond to the standard deviation.}
\label{fig:05}
\end{figure}

The reachability, $f_s$, changes substantially for the SEX and HSC networks but not as much for the other networks (Fig.~\ref{fig:05}a-d). For example, in the original SEX network about $34\%$ of the pairs of nodes were reachable in contrast to about $94\%$ in the EMA original network. After sampling, only strategy RS decreases $f_s$ in the EMA network. However, the difference with the original value is small, e.g.\ $6.4\%$ in the sampled EMA network containing about $50\%$ of the original events. This is considerably less than in the case of the SEX network that shows a difference of $55.9\%$ to the original value for the same strategy RS (Fig.~\ref{fig:05}a,b). The generally observed low biases generated by strategies NS and ES result from the redundancy of paths, i.e.\ the fact that there are multiple paths connecting the same pairs of nodes at distinct times. The absence of some events thus has little impact on $f_s$. The same redundancy is also observed for example in the SEX network but at a lesser extent, possibly because of the relatively smaller density of events in the SEX network in comparison to the EMA network (see Table~\ref{tab:01}). Furthermore, the low observed biases of strategy TS (for most data sets) indicate that the number of existing shortest paths decreases at the same rate as the number of potential paths ($N_s(N_s-1)$), for smaller $T_s$. The biases observed for SEX and HSC data sets, on the other hand, thus indicate that the new sampled nodes (introduced in the sample for increasing $T_s$) do not result in the same number of new paths as the number of potential paths that could exist (i.e.\ $f_s$ decreases with increasing $T_s$).

\begin{figure}[thb]
\centering
\includegraphics[scale=1.0]{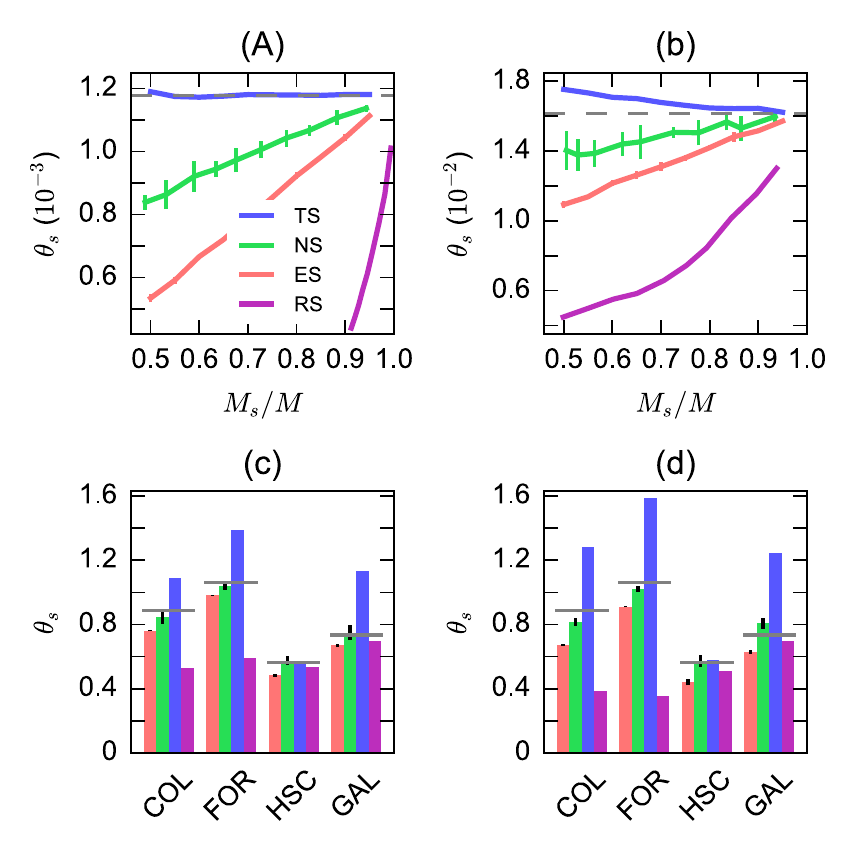}
\caption{\textbf{Temporal distance.} The average of the inverse of the temporal distance ($\theta_s$) between any pair of nodes after sampling the original network using each of the sampling strategies for the (a) SEX and (b) EMA networks. The estimation of $\theta_s$ for the COL ($\times 10^2$ hours), FOR ($\times 10^2$ days), HSC ($\times 10^3$ seconds) and GAL ($\times 10^3$ seconds) networks with configurations (c) $A$ and (d) $B$. Dashed horizontal lines correspond to the results for the original networks. Vertical bars for strategies NS and ES correspond to the standard deviation.}
\label{fig:06}
\end{figure}

Figure~\ref{fig:06}a-d shows that the statistics of the duration of the temporal paths between nodes, $\theta_s$, changes for EMA, COL, FOR and GAL for strategy TS. For the SEX and HSC data sets, this strategy generates very low biases. Although several shortest temporal paths are formed before $T_s$, some only exist if we increase $T_s$. Therefore, if we truncate the data to $T_s$, the summation term in $\theta_s$ may decrease. But since nodes are also removed (i.e.\ lower $N_s$), the overall value of $\theta_s$ increases. For the SEX and HSC data sets, the decrease in the summation term is equivalent to the decrease in the number of potential shortest paths ($N_s(N_s-1)$). On the other hand, strategy RS results in considerably different values for SEX, EMA, COL and FOR data sets. Strategy RS generates larger biases than the other strategies because higher $\delta_s$ rounds down the timings of events, collapsing many links to the same time interval and thus removing several temporal paths between nodes, that in turn results in smaller $\theta_s$. Remember that in our definition, only directly connected nodes have a temporal path within the same time step. For the other two strategies (NS and ES), uniform sampling of nodes or events increases, on average, the temporal distances between nodes. The higher $\theta_s$ given by strategy NS, in comparison to strategy ES, is possibly a result of a smaller $N_s$ obtained by strategy NS in comparison to the $N_s$ obtained by strategy ES (see Fig.~\ref{fig:02} for the SEX and EMA data sets). The relatively smaller biases in the EMA data set in comparison to the SEX data set are likely a result of higher redundancy of paths in the EMA network, as discussed in the previous paragraph.

\subsection{Epidemic Variables}

We set $\beta = 0.5$ and $\mu = 0.001$ to simulate a stochastic epidemic process. These values were chosen because they generate relatively large epidemic outbreaks in all original networks, and thus facilitate the understanding and discussion of the mechanisms regulating the epidemic process.

\begin{figure}[thb]
\centering
\includegraphics[scale=1.0]{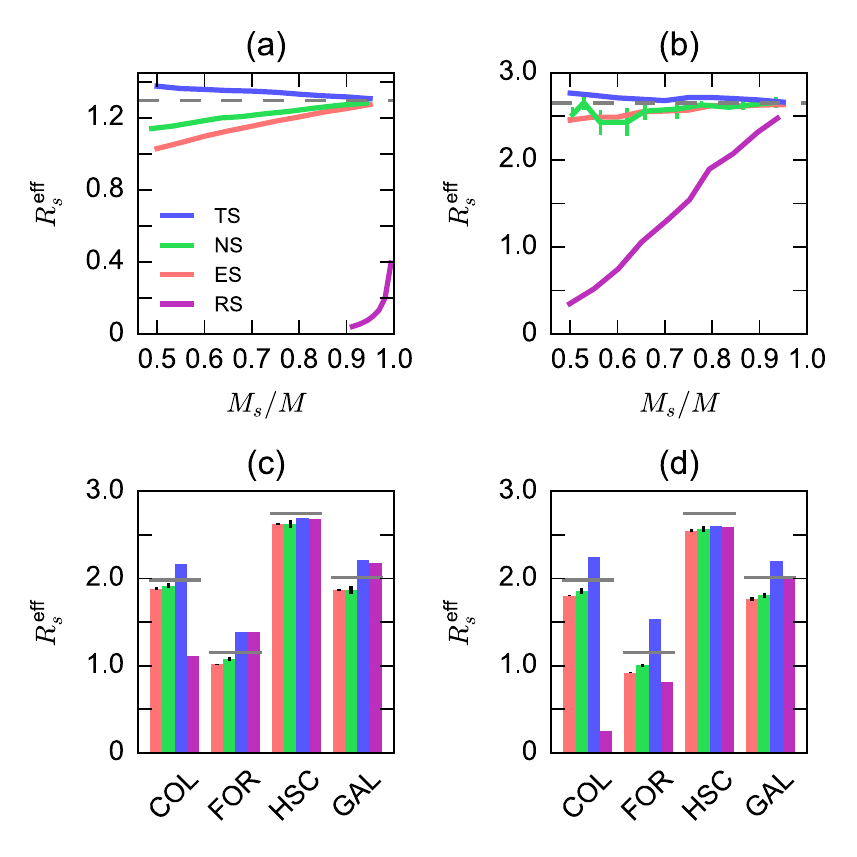}
\caption{\textbf{Secondary infections.} The average number of secondary infections ($R^{\text{eff}}_s$) after sampling the original network using each of the sampling strategies for the (a) SEX and (b) EMA networks. The estimation of $R^{\text{eff}}_s$ for the COL, FOR, HSC and GAL networks with configurations (c) $A$ and (d) $B$. Dashed horizontal lines correspond to the results for the original networks. Vertical bars for strategies NS and ES correspond to the standard deviation.}
\label{fig:07}
\end{figure}

\begin{figure}[thb]
\centering
\includegraphics[scale=1.0]{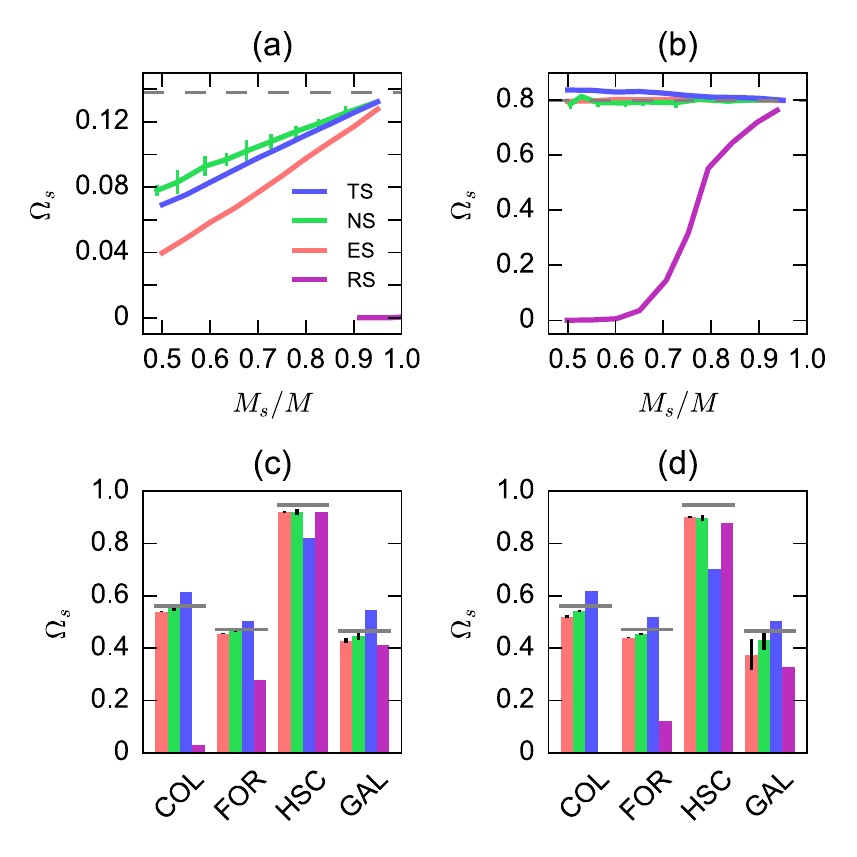}
\caption{\textbf{Outbreak size.} The average outbreak size ($\Omega_s$) after sampling the original network using each of the sampling strategies for the (a) SEX and (b) EMA networks. The estimation of $\Omega_s$ for the COL, FOR, HSC and GAL networks with configurations (c) $A$ and (d) $B$. Dashed horizontal lines correspond to the results for the original networks. Vertical bars for strategies NS and ES correspond to the standard deviation.}
\label{fig:08}
\end{figure}

We first look at the average number of secondary infections, $R^{\text{eff}}_s$. Strategy TS results in a relatively small increase in $R^{\text{eff}}_s$ for most data sets, whereas strategies NS and ES result in a small decrease for all data sets (Fig.~\ref{fig:07}a-d). The estimations of $R^{\text{eff}}_s$ given by the sampled networks indicate that the systems remain above the epidemic threshold of $R^{\text{eff}}_s = 1$ for this particular set of parameters, and that an epidemic outbreak will likely occur. Since the value of $R^{\text{eff}}_s$ also indicates how difficult is to avoid an epidemic outbreak, the estimations given by the sampled networks generally suggest that an outbreak might be easier to control than indicated by the original network (i.e.\ $R^{\text{eff}}_s$ is closer to one in the sampled networks). The results for strategy RS are substantially far from the value given by the original network for the SEX, EMA and COL data sets, but not for the FOR, HSC and GAL data sets. The low biases produced by strategies NS and ES across the different data sets are explained by the fact that the infection process is temporally finite. Many events do not actually contribute to the spread of the infection given the stochastic nature of the process, i.e.\ the absence of randomly selected interaction events has a relatively little importance to avoid infection events. The negative effect of the absence of interaction events is lower for the EMA network in which events repeat more often than in the SEX network. Therefore, the same neighbour has more chances of being infected in subsequent times in EMA than in the SEX network. This is related to the results observed for $\theta_s$ (Fig.~\ref{fig:05}) and $f_s$ (Fig.~\ref{fig:06}), where a substantial absence of events generated small biases for most networks. Strategy TS also performs well because of the finite time of the infection period that makes most infection events occur before $T_s$. If the infection period is long (small $\mu$) or the infection probability is small, the biases given by strategy TS are expected to be larger. Since the number of nodes is smaller in comparison to the original networks, $R^{\text{eff}}_s$ becomes slightly over-estimated by strategy TS. On the other hand, strategy RS generates large biases. Increasing $\delta_s$ alters the infection potential through a particular event and extends the infection period because of the rescaling of the infection and recovery probabilities, respectively. For example, in the SEX data set, if $\delta_s=7$ days, the effective infection probability is $\beta_s = \beta /7 \sim 0.07$; this infection probability is too low. Combined with the fact that the number of events (of a single node to different neighbours) at a given time step does not increase much for increasing $\delta_s$, very few neighbours may be infected by an infectious node (Fig.~\ref{fig:07}a). In the EMA network, on the other hand, there will be more events (connecting different nodes) at a single time step and thus there is a higher chance of infecting some neighbours. See also Fig.~\ref{fig:02}c for the correspondence between $M_s/M$ and $\delta_s$ for SEX and EMA data sets.

Figure~\ref{fig:08}a--d shows that the final outbreak size, $\Omega_s$, is close to zero for strategy RS applied to the SEX network, to the EMA network when approximately $65\%$ (or less) of the events are sampled, and to the COL network. For the other three sampling strategies, $\Omega_s$ is similar between the sampled and original networks for most data sets but increasingly different for smaller samples in the case of the SEX network. This is again explained by the fact that events repeat over time (less often in the SEX network). This repetition of events creates redundancies of temporal paths. In the absence of several events (by any of these three strategies), various potential infection routes remain between the nodes, and the epidemic may still grow. The biases should increase for smaller infection probabilities since an infection event will be less likely through a particular interaction event.

\section{Conclusions}


Our analyses indicate that generally both measures related to link activity are little affected by uniform sampling of nodes. This strategy also had very good performance for estimation of the statistics of temporal paths and epidemics for all network data sets but the sexual contact data set. These results likely explain the high performance of recently proposed methods to reconstruct temporal networks~\cite{Genois2015, Vestergaard2016}. That is, the temporal patterns extracted from a small sample of the temporal network are sufficient to generate larger temporal networks with realistic temporal properties. However, more research is necessary to validate these methods on diverse types of networks. Uniform sampling of events have also performed well for most statistics on most data sets. Although less efficient than uniform sampling of nodes, sampling of events may be an option when continuously collecting network data. For example, for a given number of nodes, at each time step a fraction of links may be selected and stored as time evolves (``on-line sampling''). This procedure is expected to produce better samples than truncating the observation time. In fact, truncating the observation time produced mixed results. For some networks, this sampling strategy did not affect much the statistics but for some other data sets, relatively high biases are observed (e.g.\ for lifetime and for the temporal distance). Although performing well in some cases, the poorest performance was obtained when varying the temporal resolution. In some networks, there are many repetitions of events. Therefore, merging the events on the same link by reducing the temporal resolution implies small changes in the temporal network structure. On the other hand, if there are few repetitions of events, the network might look substantially different at each temporal resolution, consequently affecting the statistics. Using a different methodology, previous research suggests that for a set of epidemiological parameters a high temporal resolution might not be necessary to study simulated epidemics in some systems~\cite{Stehle2011}.

In general, we have identified differences in the magnitude of the biases on various statistics and real-life networks. Given our results, we advice to avoid reducing much the temporal resolution but instead, if possible, we recommend uniform sampling of nodes to conserve several of the properties of temporal networks. The choice of a sampling strategy may be case-dependent, leaving some room for sampling design. In practice, it is likely to combine all proposed sampling strategies in a data collection project. It is difficult to predict the consequences of combining them since positive bias by one strategy may compensate negative bias by another strategy. Nevertheless, our study of the effects of separately applying each sampling strategy will likely improve data collection by helping the research to make informed decisions.


\bibliographystyle{unsrt}
\bibliography{main}

\end{document}